\documentclass[aps,prl,twocolumn,superscriptaddress,a4paper,showpacs]{revtex4}
\usepackage[tbtags]{amsmath}
\usepackage{amssymb}
\usepackage[utf8]{inputenc}
\usepackage{graphicx}

\newcommand{\ie}{i.e.}

\newcommand{\abs}[1]{{\left\vert #1 \right\vert}}

\newcommand{\sinch}{\mathop{\text{sinch}}}
\renewcommand{\d}{\text{d}}

\newcommand{\I}{\text{i}}
\newcommand{\e}{\text{e}}

\renewcommand{\Im}{\mathop{\text{Im}}}
\newcommand{\omc}{\omega_\mathrm{c}}
\newcommand{\isw}{{\mathrm{<}}}
\newcommand{\ilw}{{\mathrm{>}}}
\newcommand{\oms}{\omega_{L\isw}}
\newcommand{\omj}{\omega_{Lj}}
\newcommand{\omL}{\omega_L}
\newcommand{\kB}{k_\mathrm{B}}
\begin{document}
\title{Temperature dependence of Coulomb drag between finite-length quantum
  wires}
\author{J. Peguiron}
\affiliation{Department of Physics and Astronomy, University of Basel, Klingelbergstrasse 82, 4056 Basel, Switzerland}
\author{C. Bruder}
\affiliation{Department of Physics and Astronomy, University of Basel, Klingelbergstrasse 82, 4056 Basel, Switzerland}
\author{B. Trauzettel}
\affiliation{Department of Physics and Astronomy, University of Basel, Klingelbergstrasse 82, 4056 Basel, Switzerland}
\date{July 2007}
\begin{abstract}
We evaluate the Coulomb drag current in two finite-length Tomonaga-Luttinger-liquid
wires coupled by an electrostatic backscattering interaction. The drag current
in one wire shows oscillations as a function of the bias voltage applied to
the other wire, reflecting interferences of the plasmon standing waves in the
interacting wires. In agreement with this picture, the amplitude of the
current oscillations is reduced with increasing temperature. This is a clear
signature of non-Fermi-liquid physics because for coupled Fermi liquids the
drag resistance is always expected to increase as the temperature is raised.
\end{abstract}
\maketitle
%
%
Coulomb drag phenomena in coupled one-dimensional~(1D) electron systems have
been investigated quite extensively in the past 
\cite{FlePRL98,NazPRL98,KomPRL98,GurJPCM98,KlePRB00,PonPRL00,TraPRL02,PusPRL03,FiePRB06,PusPRL06}.
The interest has mainly been driven by the fact that Coulomb drag, i.e. the
electrical response of one wire as a finite bias is applied to the other
wire, seems to be an ideal testing ground for Tomonaga-Luttinger-liquid~(TLL)
physics in nature. This is because both inter-wire and intra-wire Coulomb
interactions substantially modify transport properties such as the average current and the
current noise. On the experimental side, there have been a few
works, some of which have claimed to have observed TLL behavior in the drag data 
\cite{DebPHE00,DebJPCM01,YamPHE02}. Recently, Yamamoto and coworkers have
measured Coulomb drag in coupled quantum wires of different lengths and
found peculiar transport properties that depend, for instance, on the
asymmetry of the two wires \cite{YamSCI06}. This experiment is the major
motivation for our work.

\begin{figure}[b]
  \begin{center}
  \includegraphics[width=7.5cm]{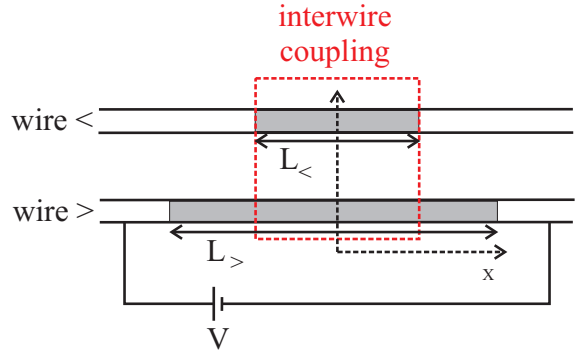}
\end{center}
  \caption{\label{fig-system}(color online). The system under consideration. Each interacting wire of length~$L_j$~[gray area, interaction parameter~$g_j(\abs{x}<L_j/2)=g_j<1$] is connected to a pair of non-interacting leads~[$g_j(\abs{x}>L_j/2)=1$]. The region of backscattering inter-wire interaction~(red dashed box) extends over the length of the shorter wire. A voltage~$V$ is applied between the leads connected to the drive wire~(here the longer wire~$j=\ilw$) and the backscattering-induced current~$I_\text{dr}$ in the drag wire~(here the shorter wire~$j=\isw$) is investigated.} 
\end{figure}

We analyze theoretically the Coulomb drag
current of two electrostatically coupled quantum wires using the concept of
the inhomogeneous TLL model \cite{MasPRB95,PonPRB95,SafPRB95}. This model is
known to capture the essential physics of an interacting 1D wire of finite
length coupled to non-interacting (Fermi liquid) electron
reservoirs. Within this framework, we are able to study finite-length and finite-temperature
effects and therefore to make qualitative
contact with the experimental setup of Ref.~\cite{YamSCI06}. Since the Coulomb
interaction varies between the wire regions and the lead regions, charge
excitations feel the interaction difference at the boundaries and are known to
exhibit Andreev-type reflections \cite{SafPRB95}. We show 
that these reflections play a crucial role in the Coulomb drag setup illustrated
in Fig.~\ref{fig-system}. Furthermore, we show that the quantum interference
phenomena associated with the Andreev-type reflections considerably modify the
drag current. This is particularly interesting as far as the temperature
dependence of the drag current is concerned. For Fermi-liquid systems, it is
well known that the drag resistance should always increase as the temperature
is raised \cite{PonPRL00}. In our setup instead, the drag current at a fixed
drive bias can either increase or decrease as a function of temperature. It
crucially depends on the interference pattern due to finite-length
effects. This is a clear signature of non-Fermi-liquid physics which could 
be observed in the double-wire setup of Ref.~\cite{YamSCI06}.

%
The system considered consists of two interacting parallel wires~($j=\isw,\ilw$) of finite
length~$L_\isw$~(shorter wire) and~$L_\ilw$~(longer wire) connected to non-interacting semi-infinite
1D leads~(Fig.~\ref{fig-system}) and is described by the
Hamiltonian $H=\sum_{j=\isw,\ilw}\left[H_j^0+H_j^V\right]+H_\mathrm{C}$.
The intra-wire interaction is modelled through a TLL description~\cite{MasPRB95,PonPRB95,SafPRB95}
\begin{equation}
  H_j^0=\frac{\hbar v_{\mathrm{F}j}}{2}\int_{-\infty}^\infty\d x\Bigl[\Pi_j^2+\frac{1}{g_j^2(x)}(\partial_x\Phi_j)^2\Bigr]
\end{equation} 
with the piecewise constant interaction parameter~$g_j(x)=g_j<1$ in the wire
region~$\abs{x}<L_j/2$ and~$g_j(x)=1$ in the non-interacting lead
regions~$\abs{x}>L_j/2$. The Fermi velocity~$v_{\mathrm{F}j}$, the interaction
strength~$g_j$, and the wire length~$L_j$ set the
frequency~$\omj=v_{\mathrm{F}j}/g_jL_j$ of the collective plasmonic
excitations hosted in each wire. A
voltage~$eV_j=\mu_\mathrm{L}^{(j)}-\mu_\mathrm{R}^{(j)}$ applied to the leads
is described by 
\begin{equation}
  H_j^V=-\frac{1}{\sqrt{\pi}}\int_{-\infty}^\infty\d x\ \mu_j(x)\partial_x\Phi_j(x),
\end{equation}
with the piecewise constant electro-chemical potential
\begin{equation}
  \mu_j(x)=\begin{cases}
    \mu_{\mathrm{L}}^{(j)} &\text{for $x<-L_j/2$},\\
    0 &\text{ for $\abs{x}<L_j/2$},\\
    \mu_{\mathrm{R}}^{(j)} &\text{for $x>L_j/2$}
  \end{cases}
  \end{equation}
with $\mu_{\mathrm{L}}^{(j)}=-\mu_{\mathrm{R}}^{(j)}$.
This model is expected to capture the essential physics of a quantum wire
coupled smoothly to electron reservoirs (with typical smoothing length $L_s$)
as long as 
$L_{j= \isw , \ilw } \gg L_s \gg \lambda_F$, where $\lambda_F$ is the electron
Fermi wavelength \cite{DolPRB05,Enss05}.
Finally, we include an inter-wire backscattering interaction over the
length~$L_\isw$ of the shorter wire, 
\begin{equation}\label{eq-defHC}
  H_\mathrm{C}=\lambda_\mathrm{BS}\int_{-L_\isw/2}^{L_\isw/2}\d x\ \cos\lbrace\sqrt{4\pi}[\Phi_\isw(x)-\Phi_\ilw(x)]\rbrace.
\end{equation}
\begin{figure}
  \begin{center}
  \includegraphics{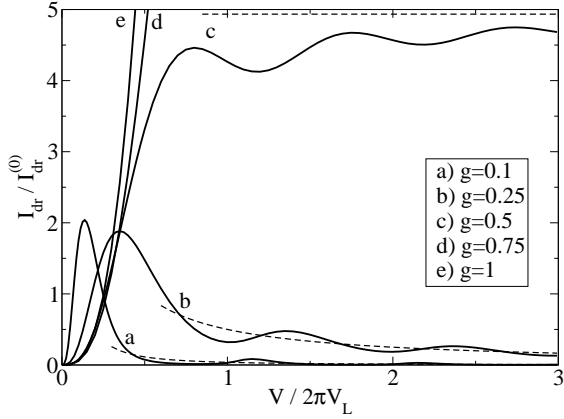}
\end{center}
\caption{\label{fig-g}Drag current as a function of drive voltage for
  identical wires at zero temperature (solid curves). 
The inter-wire interaction strength ranges from strongly interacting~($g=0.1$)
to non-interacting~($g=1$) for the different curves (with
$I_\mathrm{dr}^{(0)}=e \lambda_\mathrm{BS}^2 L^2\alpha^{4g}/\hbar^2\omL$ and~$V_L=\hbar\omL/e$). The dashed curves show the
dominant contribution~$\propto V^{4g-2}$ [given in Eq.~(\ref{eq-Idr_large_u})]
for~$g=0.1,0.25,0.5$.}  
\end{figure}
\noindent This term includes the contribution of the density-density interaction which
is most relevant to Coulomb drag~\cite{FlePRL98,NazPRL98} when the Fermi
wave-vectors of both wires are similar in magnitude,
\ie~$k_{\mathrm{F} \isw} \approx k_{\mathrm{F} \ilw}$~\footnote{Inter-wire forward-scattering
  is not explicitely included in our model as it can
  be recast in a mere renormalization of the intra-wire interaction
  strength~$g_j$ of each wire~\cite{KlePRB00}. We also 
neglect electron tunneling between
the two wires for two reasons: (i) It is a less relevant process than
$H_\mathrm{C}$ in a renormalization group sense~\cite{KomEPJ01}. (ii) It can
be tuned to zero in a Coulomb drag experiment~\cite{DebJPCM01}. We assume to
be away from half-filling where Umklapp scattering is forbidden.}.

In the following, we choose to apply a voltage~$V_\ilw=V$ to the longer
wire~($\mu_\mathrm{L}^{\ilw}=-\mu_\mathrm{R}^{\ilw}=eV/2$, drive wire) and
none~$V_\isw=0$ to the shorter
wire~($\mu_\mathrm{L}^{\isw}=\mu_\mathrm{R}^{\isw}=0$, drag wire). The average
current in the wires may then be written as $I_\isw = I_\mathrm{dr}$ and
$I_\ilw=\frac{e^2}{h}V-I_\mathrm{dr}$.
In our model, the two currents $I_\isw$ and $I_\ilw$ always flow in the 
{\it same} direction, which is due to momentum conservation. This is known as
positive Coulomb drag.
 
In order to get an
expression for the drag current~$I_\mathrm{dr}$, we follow the formalism used
in~\cite{DolPRB05} in the case of a single wire with an impurity. We consider the situation of weak inter-wire coupling. To second
order in~$\lambda_\mathrm{BS}$, we obtain
  \begin{equation}\label{eq-Idr}
    I_\mathrm{dr}=I_\mathrm{dr}^{(0)}\int_0^1\d R\int_0^{1-R}\d r j_\mathrm{dr}(r,R),
\end{equation}
with the normalization~$I_\mathrm{dr}^{(0)}={e \lambda_\mathrm{BS}^2
  L_\isw^2\alpha_\isw^{4g_\isw}}/{\hbar^2\oms}$, where~$\alpha_\isw=\oms/\omc$
is the ratio between the plasmon frequency of the shorter wire and a cutoff
frequency~$\omc$, of the order of the wire bandwidth. The plasmon frequency defines a voltage~$V_L=\hbar\oms/e$
and a temperature~$T_L=\hbar\oms/\kB$. It is convenient to introduce corresponding dimensionless voltage~$u=V/V_L$ and temperature~$\theta=T/T_L$. The integrand in~(\ref{eq-Idr}),
\begin{multline}\label{eq-jdr}
  j_\mathrm{dr}(r,R)=\frac{1}{2\alpha_\isw^{4g_\isw}}\int_{-\infty}^{\infty}\d\tau\left(\e^{\I u \tau}-\e^{-\I u \tau}\right)\\
\times\exp\left[4\pi C_\isw(r,R;\tau)+4\pi C_\ilw\left(\frac{r}{l},\frac{R}{l};\frac{q\tau}{pl}\right)\right],
\end{multline}
involves the parameters $l=L_\ilw/L_\isw$, $p=g_\ilw/g_\isw$, and $q=v_{\mathrm{F}\ilw}/v_{\mathrm{F}\isw}$. The correlation function~$C_j=C_j^\mathrm{GS}+C_j^\mathrm{TF}$ of each wire can be decomposed in a zero-temperature and a finite-temperature contribution given by~\cite{DolPRB05}
\begin{widetext}
  \begin{equation}\label{eq-defCGS}
  C_j^\mathrm{GS}(r,R;\tau)=-\frac{g_j}{4\pi}\sum_{\substack{k\in\mathbb{Z}\\ s=\pm}}\biggl[\gamma_j^\abs{2k}\ln\biggl(\frac{\alpha_j+\I(\tau-s r -2k)}{\alpha_j+\I(-2k)}\biggr)+\gamma_j^\abs{2k+1}\ln\biggl(\frac{\alpha_j+\I(\tau-s R -2k-1)}{[\alpha_j^2+(r-s R-2k-1)^2]^{1/2}}\biggr)\biggl],
\end{equation}
\begin{equation}
  C_j^\mathrm{TF}(r,R;\tau)=-\frac{g_j}{4\pi}\sum_{\substack{k\in\mathbb{Z}\\ s=\pm}}\biggl[\gamma_j^\abs{2k}\ln\biggl(\frac{\sinch[\pi\theta(\tau-s r-2k)]}{\sinch[\pi\theta(-2k)]}\biggr)
+\gamma_j^\abs{2k+1}\ln\biggl(\frac{\sinch[\pi\theta(\tau-s R-2k-1)]}{\sinch[\pi\theta(r-s R- 2k-1)]}\biggr)\biggr],
\end{equation}
\end{widetext}
with~$\gamma_j=(1-g_j)/(1+g_j)$ and~$\sinch x = (\sinh x)/x$. 

It is to be noted that the expression for the drag current~$I_\mathrm{dr}$ does not depend on whether the drive wire is the longer wire or the shorter one due to the symmetry of our model.
In the following, we present results obtained by numerical evaluation of the
triple integral involved in~(\ref{eq-Idr}) and~(\ref{eq-jdr}) and discuss
several analytical approximations.

First we set the temperature to zero and consider identical wires ($l=p=q=1$, thus we drop the wire index~$j$).
The drag current shows non-monotonous behavior and oscillations with period~$\sim 2\pi\hbar\omL/e$ as a function of the bias voltage~(Fig.~\ref{fig-g}).
It decays at large voltages for~$g<1/2$ whereas it increases for~$g>1/2$.
Thus, we obtain qualitatively the same behavior as in a dual Coulomb drag
setup where a drive current is applied and a drag voltage is
measured~\cite{NazPRL98}. Similar oscillations as a function of voltage have
been predicted in the context of two coupled fractional quantum Hall line
junctions~\cite{zuel04}.

An analytic approximation can be derived in the limit~$u=eV/\hbar\omL\gg1$, that is for high voltages or long wires.
In Eq.~(\ref{eq-defCGS}), the terms proportional to~$\gamma^{\abs{m}}$ account for contributions from plasmon excitations reflected $\abs{m}$~times inside the wire.
When the wire length is much longer than other relevant length scales, the contribution without any reflection~$m=0$ becomes dominant and yields the integrand
\begin{equation}\label{eq-jdr_m0}
  j_\mathrm{dr}(r,R)\underset{u\gg1}{\sim}\frac{\pi^{3/2}}{\Gamma(2g)}\left(\frac{u}{2r}\right)^{2g-1/2} J_{2g-1/2}(ur),
\end{equation}
where~$\Gamma(z)$ denotes the Gamma function and~$J_\nu(z)$ the 
Bessel function of order~$\nu$.
The resulting expression for the drag current, which involves hypergeometric functions, underestimates the amplitude of the oscillations with respect to the exact numerical result.
However, the behavior at large~$u$, governed by the dominant contribution 
\begin{equation}\label{eq-Idr_large_u}
  I_\mathrm{dr}\underset{u\gg1}{\sim}I_\mathrm{dr}^{(0)}\frac{\pi^2}{2 \Gamma^2(2g)}\left(\frac{u}{2}\right)^{4g-2},
\end{equation}
shows good agreement in the appropriate parameter regime~(dashed curves in
Fig.~\ref{fig-g}). Since we do perturbation theory in $\lambda_{\rm BS}$,
the relation $I_\mathrm{dr} \ll (e^2/h)V$ has to hold. In the
large $u$ regime, this means $(\lambda_{\rm BS} L/\hbar
\omega_c)^{2} (\omega_L/\omega_c) (eV/\hbar \omega_c)^{4g-3} \ll 1$.



\begin{figure}
  \begin{center}
  \includegraphics{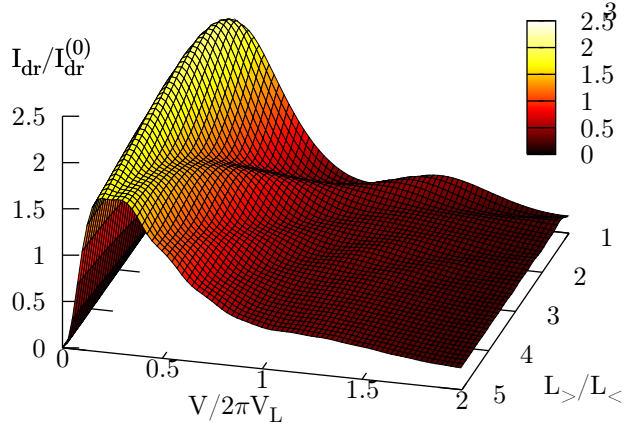}
  \end{center}
  \caption{\label{fig-l}(color online). Drag current as a function of the bias
    voltage and of the length ratio of the wires (with $I_\mathrm{dr}^{(0)}=e \lambda_\mathrm{BS}^2 L_\isw^2\alpha_\isw^{4g}/\hbar^2\oms$ and~$V_L=\hbar\oms/e$).}
\end{figure}

Now we consider the situation where the two wires have different lengths.
The qualitative behavior of the drag current does not change for increasing length ratio~$l=L_\ilw/L_\isw$, but the peak positions get shifted to lower voltages~(Fig.~\ref{fig-l}).
Here, neglecting plasmon reflections in the correlation function of the wires leads again to the expression~(\ref{eq-jdr_m0}), which is independent of~$l$.
This fact explains why the drag current does not change appreciably
as a function of~$l$ and indicates that the peak shifts observed result from
plasmon reflections inside the wires. Our studies are the first to analyze the
effect of an asymmetry in the length on Coulomb drag phenomena in coupled
quantum wires which is of recent experimental
relevance~\cite{YamSCI06}. 



\begin{figure}
  \begin{center}
    \includegraphics{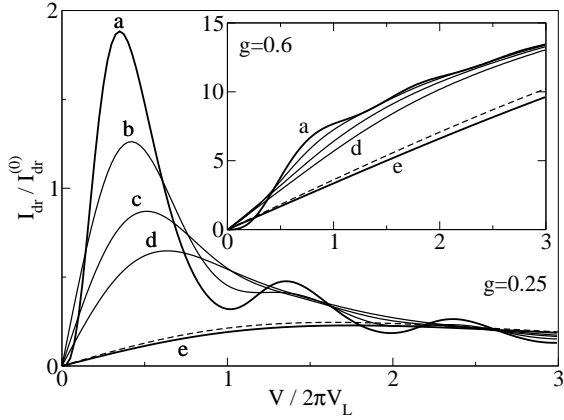}
  \end{center}
    \caption{\label{fig-T}Temperature dependence of the drag current for identical wires with interaction strength~$g=0.25$.
The solid curves (labelled a - e) are evaluated for temperatures given by
$T/T_L  = \kB T / \hbar \omL = 0, 0.5, 1, 1.5, 5$, respectively (with $I_\mathrm{dr}^{(0)}=e \lambda_\mathrm{BS}^2 L^2\alpha^{4g}/\hbar^2\omL$ and~$V_L=\hbar\omL/e$).
The dashed curve shows the high-temperature limit~(\ref{eq-Idr_large_theta})
for~$T/T_L=5$. The inset shows a similar plot for weaker intra-wire
  interaction $g=0.6$ where the oscillations are less pronounced.} 
\end{figure}

We now discuss in detail the temperature
dependence of the drag current. For clarity, we consider
again symmetric wires ($l=1$)~\footnote{In view of the results shown in Fig.~\ref{fig-l}, we do not expect any qualitative changes of the predicted
temperature dependence for the case $l=1$ as we make the wires asymmetric in
length ($l \neq 1$).}. 
The oscillations of the drag current as a function of bias
voltage get washed out with increasing temperature~(Fig.~\ref{fig-T}). 
This behavior is consistent with the picture which attributes the oscillations to interferences of the plasmon excitations of the wires.
Thus, for bias voltages such that the drag current is close to a
maximum at zero temperature, one observes a
decrease of the drag current with increasing temperature, whereas
the opposite behavior can be observed close to a minimum of the
zero-temperature drag current for strong
interactions.  This behavior is in stark contrast to the linear temperature
dependence predicted for Coulomb drag between 1D Fermi-liquid
conductors~\cite{GurJPCM98} and therefore a clear signature of TLL
physics. Note that our Fig.~\ref{fig-T} bears significant resemblance to
Fig.~9 of Ref.~\cite{DebJPCM01}.

A good approximation of the drag current can be obtained for temperatures much
larger than the temperature associated with the plasmon frequency,~$\theta=\kB
T/\hbar\omL\gg1$, by neglecting contributions from plasmon reflections in the
wires. Then, we obtain the dominant contribution
\begin{equation}\label{eq-Idr_large_theta}
  I_\mathrm{dr}\underset{\theta\gg1}{\sim}I_\mathrm{dr}^{(0)}(2\pi\theta)^{4g-2}\sinh\left(\frac{u}{2\theta}\right)\frac{\abs{\Gamma\left(g+\frac{\I u}{4\pi\theta}\right)}^4}{4\Gamma^2(2g)}.
\end{equation}
Taking the limit of low bias voltage in this result,
\begin{equation}
  I_\mathrm{dr}\underset{\theta\gg1,u}{\sim}I_\mathrm{dr}^{(0)}\frac{\pi \Gamma^4(g)}{4\Gamma^2(2g)}(2\pi\theta)^{4g-3} u, 
\end{equation}
we recover the power-law dependence~$T^{4g-3}$ of the linear conductance predicted by renormalization group analysis~\cite{KlePRB00}.
At large bias voltage~$u\gg\theta\gg1$, we recover the zero-temperature result~(\ref{eq-Idr_large_u}).

\begin{figure}[t!]
  \begin{center}
    \includegraphics{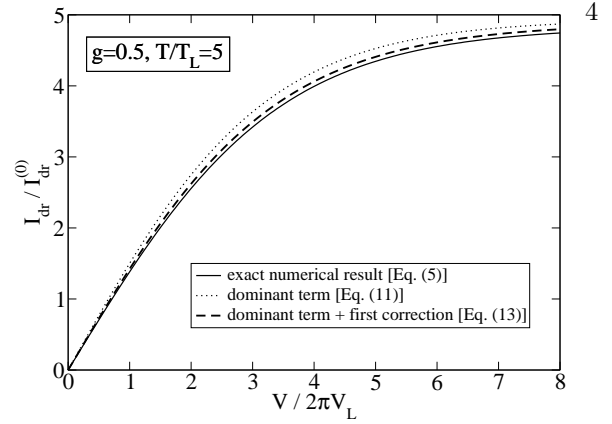}
  \end{center}
    \caption{\label{fig-Tg05}Drag current at high temperature~($T/T_L=5$) for
      identical wires with interaction strength~$g=0.5$~(solid curve), where
      $I_\mathrm{dr}^{(0)}=e \lambda_\mathrm{BS}^2 L^2\alpha^{4g}/\hbar^2\omL$
      and~$V_L=\hbar\omL/e$. 
The dotted curve shows the dominant contribution in the high-temperature limit~[first term in~(\ref{eq-Idr_large_theta_g05})], the dashed curve includes the first correction as well~[both terms in~(\ref{eq-Idr_large_theta_g05})].} 
\end{figure}

In the case~$g=1/2$, Eq.~(\ref{eq-Idr_large_theta}) as well as the contribution to next order in~$\theta^{-1}$ can be brought into a compact analytic form
\begin{multline}\label{eq-Idr_large_theta_g05}
  I_\mathrm{dr}\underset{\theta\gg1}{\sim}I_\mathrm{dr}^{(0)}\biggl[\frac{\pi^2}{2}\tanh\left(\frac{u}{4\theta}\right)\\
  +\frac{1}{4\theta}\Im\left\lbrace\Psi^\prime\left(\frac{1}{2}+\frac{\I u}{4\pi\theta}\right)\right\rbrace\biggr],
\end{multline}
where~$\Psi^\prime(z)=\d^2\ln\Gamma(z)/\d z^2$ denotes the trigamma function.
The first term is Eq.~(\ref{eq-Idr_large_theta}) evaluated at~$g=1/2$~(dotted
curve in Fig.~\ref{fig-Tg05}), and the second one is the dominant correction,
which takes values within~$[-I_\mathrm{dr}^{(0)}/\theta,0]$~(included in the
dashed curve in Fig.~\ref{fig-Tg05}).
This illustrates that the first-order approximation already yields a nice
qualitative description for the full numerical result, which makes us confident that Eq.~(\ref{eq-Idr_large_theta}) is a good high-temperature approximation also for~$g\neq 1/2$.

In summary, we have analyzed two coupled quantum wires that exhibit both a
finite intra-wire interaction and a finite inter-wire interaction. We have
taken into account finite-length effects within the inhomogeneous TLL model
that is known to capture the essential physics of quantum wires coupled to
Fermi-liquid reservoirs. We have investigated how an asymmetry in the lengths
of the wires changes the drag current and we have
predicted a rich temperature dependence of the drag current that shows 
clear signatures of non-Fermi-liquid physics.

We would like to thank F.~Dolcini, H.~Grabert, M.~Kindermann, Y.~Nazarov, S.~Tarucha, and M.~Yamamoto
for interesting discussions. This work was supported by the Swiss NSF and the
NCCR Nanoscience.

\end{document}